\documentclass[english,notitlepage,superscriptaddress,nofootinbib,twocolumn]{revtex4-2}
\usepackage[T1]{fontenc}
\usepackage[utf8]{inputenc}
\usepackage{amsmath}
\usepackage{amssymb}
\usepackage{graphicx}
\usepackage{braket}
\usepackage[english]{babel}
\usepackage{hyperref}
\hypersetup{colorlinks=true,urlcolor=blue,citecolor = blue}
\usepackage{multirow}

\usepackage{nicefrac}
\usepackage{float}
\usepackage{siunitx}
\usepackage{bbold}

\usepackage[newcommands]{ragged2e}
\usepackage{graphicx,caption}
\usepackage{subfigure}
\usepackage[font=small,labelfont=bf,justification=justified]{caption}

\newcommand{\Max}[1]{\raisebox{0.5ex}{\scalebox{0.8}{$\displaystyle \max_{#1}\;$}}}

\begin{document}
\title{Relativistic Effects on Entangled Single-Electron Traps}

\newcommand{\affone}{University College London, Gower Street, WC1E 6BT London, United Kingdom}
\newcommand{\afftwo}{Physics Institute, University of São Paulo, Rua do Matão 1371, São Paulo, Brazil}
\newcommand{\affthree}{Van Swinderen Institute, University of Groningen, 9747 AG, The Netherlands}

\author{Marko Toro\v{s}}
\affiliation{\affone}
\author{Patrick Andriolo}
\affiliation{\afftwo}
\author{Martine Schut}
\affiliation{\affthree}
\author{Sougato Bose}
\affiliation{\affone}
\author{Anupam Mazumdar}
\affiliation{\affthree}

\newcommand{\mt}[1]{{\color{blue}MT:''#1''}}

\begin{abstract}
{
The manipulation of individual charged particles has been deeply explored in physics's theoretical and experimental domains during the past decades. It is the pillar of several existing devices used for metrology and sensing and is a promising platform for realizing future technologies, such as quantum computers. It is also known that in the relativistic regime, interactions between charged particles
become affected by post-Coulombian corrections, with the dominant couplings encoded in the \textit{Darwin Hamiltonian}.
The Darwin term has been extensively studied in atomic physics, where the interaction range is confined to the sub-angstrom scale. Still, there is a lack of understanding about whether (and when) Darwin's contributions are relevant at larger scales.
In this paper, we explore the effects of these corrections in a system of two harmonically trapped electrons, 
where we look into the behaviour of quantum entanglement present in the static and dynamical regimes. We explore the parameter space of the developed model and seek frequencies, distances, and squeezing parameters
for which relativistic effects become relevant for the generation of entanglement.}
\end{abstract}

\maketitle

\section{Introduction}



Trapping and manipulation of quantum systems has undoubtedly been one of the most successful areas of quantum science~\cite{kastner1997technology,blatt2008entangled,blinov2004quantum,friis2018observation}. Charged particles, such as ions or electrons, can be stably trapped using time-dependent electric fields in Paul traps~\cite{RevModPhys.62.531,leibfried2003quantum}, evading Earnshaw's theorem of classical electromagnetism~\cite{jackson1999classical}. Such systems have already found numerous applications in precision measurements~\cite{thompson1990precision} and in metrology~\cite{rosenband2008frequency}.
 
The trapping of charged particles also paves the way for many novel applications, ranging from tests of fundamental physics \cite{bushev2016single,quint2014fundamental,carney2021trapped,haustein2019mass} to the possibility of designing scalable quantum computers \cite{kielpinski2002architecture,monroe1995demonstration,yu2022feasibility,marzoli2009experimental,ciaramicoli2004trapped}. In particular, single-electron traps 
offer the possibility of using the individual spin of these particles as a resource for manipulating quantum information~\cite{wineland1973monoelectron,brown1986geonium,jefferson2002two,yamahata2014gigahertz,segal2006nanoscale,daniilidis2013quantum}.

Many experimental groups have demonstrated trapping of electrons in recent decades~\cite{yamahata2014gigahertz,segal2006nanoscale,matthiesen2021trapping,peng2017spin,lientschnig2002single,ciaramicoli2003scalable}, which consists of maintaining the electrons in a confined region through modulated electromagnetic fields. The control of the electron's motional state as well as of its spin state offers a versatile system to explore quantum phenomena. 
In addition, the experimental control of single-electron traps may enable probing of the regime where relativistic effects become relevant~\cite{Sonnleitner_2018,Schwartz_2019}.

In this article, we show that, under certain conditions, relativistic effects can present significant corrections to the usual Coulomb interaction between two electrons in adjacent traps. The first post-Coulombian (PC) correction is encoded in the famous Darwin Hamiltonian, which can be derived directly from quantum electrodynamics~\cite{landau-lifshitz-course,bethe2013quantum}.
Quantum systems that interact electromagnetically tend to become entangled~\cite{yang2015electromagnetically,Horodecki_2009}, and we will explore how the presence of relativistic corrections changes the build-up of quantum correlations.

Recently there has been much interest in probing the fundamental nature of physics in table-top experiments via entanglement, such as conceiving
a protocol to test quantum properties of gravity using two
massive quantum particles in quantum superpositions~\cite{Bose:2017nin,marletto2017gravitationally}~\footnote{The results of Ref.~\cite{Bose:2017nin} were already known earlier,
see~\cite{ICTS}.}. This protocol, known as the quantum gravity-induced entanglement of masses (QGEM), probes the quantum nature of gravitational perturbations around the weak Minkowski background~\cite{Bose:2017nin,Marshman:2019sne,Bose_2022,Carney_2019,Christodoulou:2022vte,christodoulou2019possibility,Belenchia:2018szb,Danielson:2021egj,Biswas:2022qto,Vinckers:2023grv,Elahi:2023ozf, Hanif:2023fto}. In this regard,  QGEM is similar to Bell's original idea of
testing quantum correlations (i.e., Bell correlations) between two
spatially separated systems, see~\cite{bell1964einstein,brunner2014bell}. Moreover, entanglement-based tests have been proposed to study quantum correlations between matter and photon~\cite{Biswas:2022qto,Carney:2021vvt}. As highlighted in \cite{Biswas:2022qto}, this is the quantum version of the classical light bending experiment, which can distinguish the spin-2 nature of the gravitational interaction, e.g. massless spin-2 graviton as a quantum mediator. Furthermore, one can also probe the dark matter-induced entanglement~\cite{Barker:2022mdz}, test the quantum version of the weak equivalence principle~\cite{Bose:2022czr, Chakraborty:2023kel}, test a massive graviton~\cite{Elahi:2023ozf}, and modified theories of gravity motivated from string theory~\cite{Vinckers:2023grv}. Moreover, we recently considered post-Newtonian corrections to the gravitational entanglement, showing that relativistic corrections can strongly suppress quantum correlations for certain parameter spaces, 
while classically the gravitational Hamiltonian remains attractive~\cite{toros2024relativistic}. 

This paper is organized as follows. Sec.~\ref{sec:harmonic-traps} introduces the quantum modelling of the two adjacent traps. Sec.~\ref{sec:darwin-static} details how static entanglement is formed when relativistic corrections are taken into account. In Sec.~\ref{sec:darwin-dynamic}, we extend our analysis to the formation of dynamical entanglement by using squeezed states. Sec.~\ref{sec:conclusion} exhibits our concluding remarks, while Appendix~\ref{app:A} presents the higher-order corrections of the Darwin Hamiltonian and Appendix~\ref{app:B} provides details about the formalism used for treating continuous bosonic systems employed throughout this article.


\section{Charged particles in Harmonic Traps with Post-Coulombian corrections}\label{sec:harmonic-traps}

Our model consists of two particles trapped in two adjacent harmonic traps on the x-axis and centred at $\pm d/2$. We assume that each trap contains a single particle of mass $m$ and unit charge $-e$ (i.e., two harmonically trapped electrons). We furthermore suppose that the mechanical frequency of the two traps is the same and will be denoted by $\omega$. As we will see, this simplified one-dimensional model is sufficient to explore the parameters space of post-Coulombian corrections to the entangling power of electromagnetic interactions. At the same time, we leave detailed feasibility studies for future research.

The positions of the two particles can be written as 
\begin{equation}
\hat{r}_1 = -d/2 + \hat{x}_1, \quad \hat{r}_2 = d/2 +  \hat{x}_2,
\end{equation}
for the left and right particles, respectively. The fluctuations around the equilibrium positions are denoted by $\hat{x}_1$ and $\hat{x}_2$ and the corresponding conjugate momenta are given by $\hat{p}_1$ and $\hat{p}_2$. The matter Hamiltonian is, in this case, given by
\begin{align}
    \hat{H} = \sum_{i=1,2} \left(\frac{\hat{p}_i^2}{2m} + \frac{1}{2}m\omega^2 \hat{x}_i^2\right) + \hat{H}_\text{int},
\end{align}
where $\hat{H}_\text{int}$ contains the electromagnetic interactions up to the first post-Coulombian (PC) terms, i.e., up to order $\mathcal{O}(c^{-2})$, where $c$ denotes the speed of light. Specifically, the considered interaction consists of two terms:
\begin{align}
    \hat{H}_\text{int} = \hat{H}_\text{C} + \hat{H}_\text{D},
\end{align}
with $\hat{H}_\text{C}$ being the usual electrostatic Coulomb potential (i.e., the 0PC term), and $\hat{H}_\text{D}$ corresponding to the Darwin Hamiltonian (i.e., the 1PC term). In particular, $\hat{H}_\text{D}$ is composed of momentum contributions that arise at the leading order relativistic correction. From now on, the subscripts ``C'' and ``D'' will refer to quantities associated with Coulomb and Darwin terms, respectively. The form of these terms in the interaction Hamiltonian is given by~\footnote{Note that here we are not discussing the vacuum entanglement in quantum electrodynamics; for such a description, see~\cite{Fedida:2024dwc}. }:
\begin{alignat}{2}
    \hat{H}_\text{C} &= \frac{e^2}{4 \pi \varepsilon_0 \vert\hat{r}\vert}, \label{Coulumb} \\
     \hat{H}_\text{D} &= -\frac{e^2}{8 \pi \varepsilon_0 m^2 c^2} \left[\frac{\left({\hat{p}}_1 \cdot {\hat{p}}_2\right)}{\vert\hat{r}\vert}
     + \frac{\left({\hat{p}}_1 \cdot \hat{r}\right)\left({\hat{p}}_2 \cdot \hat{r}\right)}{\vert\hat{r}\vert^3}\right], \label{eq:hamiltonian-darwin}
\end{alignat}
where the distance between the two particles is given by
\begin{equation}
    \hat{r}\equiv \hat{r}_2 - \hat{r}_1= d +\hat{x}_2 - \hat{x}_1. \label{rx}
\end{equation} 
For the derivation of Eqs.~\eqref{Coulumb} and \eqref{eq:hamiltonian-darwin}  starting from quantum electrodynamics, see, for example, chapter 9 in Landau - Lifschitz~\cite{landau-lifshitz-course} (see also alternative derivations in~\cite{Schwartz_2019,Sonnleitner_2018}). While here we limit the analysis to 0PC and 1PC terms for simplicity, higher-order corrections could, in principle, be obtained from the Breit Hamiltonian and analyzed in an analogous way~\cite{bethe2013quantum}. 

We now decompose the position and momenta fluctuations in terms of the ladder operators,
\begin{equation}
    \hat{x}_{j} = \delta x \left(\hat{a}_j^{\dagger}+\hat{a}_j\right),\quad
\hat{p}_{j}= i \delta p \left(\hat{a}_j^{\dagger}-\hat{a}_j\right), \label{ladder}
\end{equation}
with $j=1,2$. The zero-point-motions are given by
\begin{equation}\label{zpf}
    \delta x \equiv  \sqrt{\hbar/(2 m \omega)} \quad \delta p \equiv  \sqrt{ m \omega \hbar/ 2},
\end{equation}
where we recall that for simplicity, we assume that the mass and mechanical frequency are the same for the two quantum harmonic oscillators. We can now insert Eq.~\eqref{rx} into Eqs.~\eqref{Coulumb} and \eqref{eq:hamiltonian-darwin} to obtain the leading order interaction terms between the two particles by series expanding in $\hat{x}_1$ and $\hat{x}_2$ (i.e., we assume that each particle is close to the centre of the trap such that the distance $d$ between the two traps is much larger compared to the spatial delocalization of the particles). 

By further using the decomposition in Eq.~\eqref{ladder}, we can classify the interaction terms between the two particles in terms of powers of $\hbar$ (we note that $\delta x$ and $\delta p$ are both proportional to $\sqrt{\hbar}$). The position-position coupling $\propto \hat{x}_1 \hat{x}_2$ from the Coulomb interaction, and the momentum-momentum coupling $\propto \hat{p}_1 \hat{p}_2$ from the Darwin interaction, provide the leading order interaction between the two particles at order $\mathcal{O}(\hbar)$ (for a discussion of higher order corrections see Appendix~\ref{app:A}). 
The corresponding interaction Hamiltonian is given by
\begin{align}\label{hamiltonianequation}
\begin{aligned}
    \hat{H}_\text{int} &= \hbar (g_\text{C} - g_\text{D})  (\hat{a}_1^\dagger \hat{a}_2 + \hat{a}_1 \hat{a}_2^\dagger)\\
    &\hspace{0.5cm}+ \hbar (g_\text{C} + g_\text{D}) (\hat{a}_1 \hat{a}_2 + \hat{a}_1^\dagger \hat{a}_2^\dagger),
\end{aligned}
\end{align}
where the couplings arising from the Coulomb and Darwin Hamiltonian have the following expressions
\begin{align}\label{couplings}
    g_\text{C}\equiv -\frac{e^2}{4\pi \varepsilon_0 m\omega d^3} \ , \ \  g_\text{D}\equiv \frac{3\omega e^2}{16\pi\varepsilon_0 m d c^2},
\end{align}
respectively. The corresponding quantum phases, which can be used to make an order of magnitude estimates, are given by $\Delta\phi_C=g_\text{C}\tau$ and $\Delta\phi_D=g_\text{D}\tau$, respectively, where $\tau$ denotes the time
\footnote{The induced phase by the Darwin interaction is explicitly given by
    $\Delta \phi_\text{D} = \frac{3\omega e^2 \tau}{16\pi \varepsilon_0 m dc^2}$,
so the required coherence time scales as $\tau \propto d/\omega$, which suggests that experiments with smaller distances and higher frequencies would be advantageous. If we seek to find an induced phase of $\Delta \phi_\text{D}\sim {\cal O}(1)$, for a pair of electrons in two adjacent traps of frequency $\omega\sim 1$ GHz~\cite{yamahata2014gigahertz,glasson2005trapping,matthiesen2021trapping} and located at distances $d$ of $250$nm, $500$nm and $750$nm, we find that the required coherence time $\tau$ is about $0.35$s, $0.7$s and $1$s, respectively. While here we do not enter into detailed experimental feasibility considerations, we point the interested reader to single-electron traps~\cite{brown1986geonium,segal2006nanoscale,kastner1997technology,bushev2016single,wineland1973monoelectron,matthiesen2021trapping,lientschnig2002single,yu2022feasibility,fan2022improved}.
\label{firstfootnote}}.

In the following sections, we will explore the consequences of the Hamiltonian in Eq.~\eqref{hamiltonianequation}. We note that $g_\text{C}$ is a 0PC coupling (i.e., non-relativistic), while $g_\text{D}$ is a 1PC coupling (i.e., first relativistic correction at order $\mathcal{O}(c^{-2})$), with the two couplings having opposite sign. We will explore the consequences of these couplings first for energy eigenstates, offering a glimpse into the phenomenology (Sec.~\ref{sec:darwin-static}), and then using squeezed states, where we can control the delocalizations by tuning the degree of initial squeezing (Sec.~\ref{sec:darwin-dynamic}).


\section{Static Darwin Effects}\label{sec:darwin-static}
Quantum systems that evolve under nontrivial dynamics will generally become entangled; see for a review~\cite{Horodecki_2009}. In particular, the system composed of the two harmonically trapped particles becomes entangled with our interaction Hamiltonian acting as a perturbation~\cite{Bose:2017nin,Marshman:2019sne, Bose_2022,toros2024relativistic}. 
We will assume that the system is initialized in the global vacuum state of the two harmonic oscillators, $\ket{00}\equiv \vert0\rangle\vert0\rangle$,  which will allow us to obtain simple formulae using first-order non-degenerate perturbation theory. The final state of our bipartite system can be written on a number basis as
\begin{align}
    |\psi \rangle = \frac{1}{\sqrt{\mathcal{N}}}\sum_{m,n} \lambda_{mn}\ket{mn} 
    \label{eq:final-state-def}
\end{align}
where $\mathcal{N} = \sum_{m,n}|\lambda_{mn}|^2$ is the normalization factor and $\lambda_{mn}$ are the coefficients corresponding to the state $\ket{mn}\equiv \vert m \rangle \vert n \rangle$ (with $\vert m \rangle$, $\vert n \rangle$ denoting number states). 
The unperturbed state is associated to the coefficient $\lambda_{00}=1$, and the other coefficients  can be computed using
\begin{align}
    \lambda_{mn}= \frac{\langle m n|\hat{H}_\text{int}| 0 0\rangle}{2 E_0-E_m-E_n},\label{eq:coeff-def}
\end{align}
where $E_j$ is the energy associated to the number state $\vert j\rangle $ (with $j\in \mathbb{N}$).

From Eqs.~\eqref{hamiltonianequation} and \eqref{eq:coeff-def} we see that the only terms of the interaction Hamiltonian  $\hat{H}_\text{int}$ that contribute are given by 
 \begin{align}
   \hat{H}_\text{int} \propto \hbar (g_\text{C}+g_\text{D}) \hat{a}_1^\dagger \hat{a}_2^\dagger 
    =   \frac{\hbar e^2}{4\pi\varepsilon_0 m d}\left(\frac{3\omega}{4c^2} - \frac{1}{\omega d^2}\right) \hat{a}_1^\dagger \hat{a}_2^\dagger.\label{eq:ham-static}
\end{align}
Specifically, when we apply $\hat{a}_1^\dagger \hat{a}_2^\dagger$ to the initial state $\ket{00}$ we generate the state $\ket{11}$. According to Eqs.~\eqref{eq:final-state-def}-\eqref{eq:ham-static} the amplitude of the state  $\ket{11}$ is controlled by competition of two parameters: by the mechanical frequency, $\omega$,  and by the distance between the two traps, $d$. 
For arbitrarily small frequencies, the Coulomb term dominates, but after the frequency exceeds a threshold $\omega = {2c}/({\sqrt{3}d})$, the relativistic effects overpasses the electrostatic interaction.

To summarize, using Eqs.~\eqref{eq:final-state-def}-\eqref{eq:ham-static}, we find that the perturbed state is given by:
\begin{align}\label{eq:9}
    \ket{\psi} = \frac{1}{\sqrt{1+\lambda^2}}\left(\vert 00\rangle + \lambda\vert 11\rangle\right),
\end{align}
with the coupling $ \lambda \equiv  \lambda_{11}$ expressed as:
\begin{align}\label{lambdacoef}
    \lambda = \frac{e^2}{8\pi \varepsilon_0 m \omega^2 d^3} \left(1-\frac{3 d^2 \omega^2}{4c^2}\right).
\end{align}
In the non-relativistic limit, we recover the result that one would expect by considering only the Coulomb interaction, i.e., 
\begin{align}\label{limit}
     \lambda_\text{NR}=\lim_{c\rightarrow \infty}\lambda = \frac{e^2}{8\pi \varepsilon_0 m \omega^2 d^3}.
\end{align}
Let us make some initial observations. Since Eq. (\ref{eq:9}) represents a two-qubit state, its separability can be understood by looking at the behaviour of the coefficient $\lambda(d,\omega)$. $\ket{\psi}$ remains separable if and only if $\lambda=0$, which occurs when $\omega = {2c}/{\sqrt{3}d}$. From Eq.~\eqref{lambdacoef} we note that the relativistic correction decays slower with the distance in comparison with the Coulomb interaction, i.e., for every frequency $\omega$ there is a distance $d$ beyond which relativistic corrections become more pronounced than the Coulomb interaction. If we work in the non-relativistic limit by taking the limit $c\rightarrow\infty$ in Eq.~\eqref{limit} the cancellation point disappears. In other words, the separability of the state $\ket{\psi}$ results from a non-trivial cancellation between 0PC Coulomb and 1PC Darwin contributions.

Going beyond the coefficient analysis, we can explore the quantum correlations in more detail with an entanglement measure. Here, we will quantify the generated entanglement using the von Neumann entanglement entropy~\cite{Horodecki_2009}.
Given sets of basis vectors $\ket{u_i}$ and $\ket{v_i}$ for the first and second subsystem, respectively, the Schmidt decomposition of a general bipartite state $\ket{\phi}$ is given by
\begin{align}
    \ket{\phi} = \sum_{i} \lambda_i \ket{u_i v_i}, \label{decomposition}
\end{align}
where $\lambda_i$ are called Schmidt coefficients. The associated von Neumann entanglement entropy is given by
\begin{align}
    S(\ket{\phi}) = -\sum_{i}\lambda_i^2\log_2(\lambda_i^2).\label{eq:VNE}
\end{align}
In our case, the state (Eq.~\eqref{eq:9}) is already in the form of Eq.~\eqref{decomposition} and we can readily extract the Schmidt coefficients, i.e., $1/\sqrt{1+\lambda^2}$ and $\lambda/\sqrt{1+\lambda^2}$. Taking into account that $\lambda \ll 1$ we then find that the von Neumann entanglement entropy of the perturbed state in eq.~\eqref{eq:9} is given by a simple expression:
\begin{align}\label{entanglemententropyexpression}
     S(\ket{\psi}) = -\lambda^2\log(\lambda^2),
\end{align}
where $\lambda$ has been defined in Eq.~\eqref{lambdacoef}.
%
Eq.~\eqref{entanglemententropyexpression} allows us to readily compute the entanglement entropy $S_{C+D}$ arising from the Coulomb and Darwin terms (i.e., both the 0PC and 1PC terms) and the entanglement entropy $S_{C}$ corresponding to the non-relativistic limit (i.e., the 0PC term): we insert the coefficients from Eqs.~\eqref{lambdacoef} and \eqref{limit} into Eq.~\eqref{entanglemententropyexpression}, respectively. 
\begin{figure}[t]
        \includegraphics[width=1\linewidth]{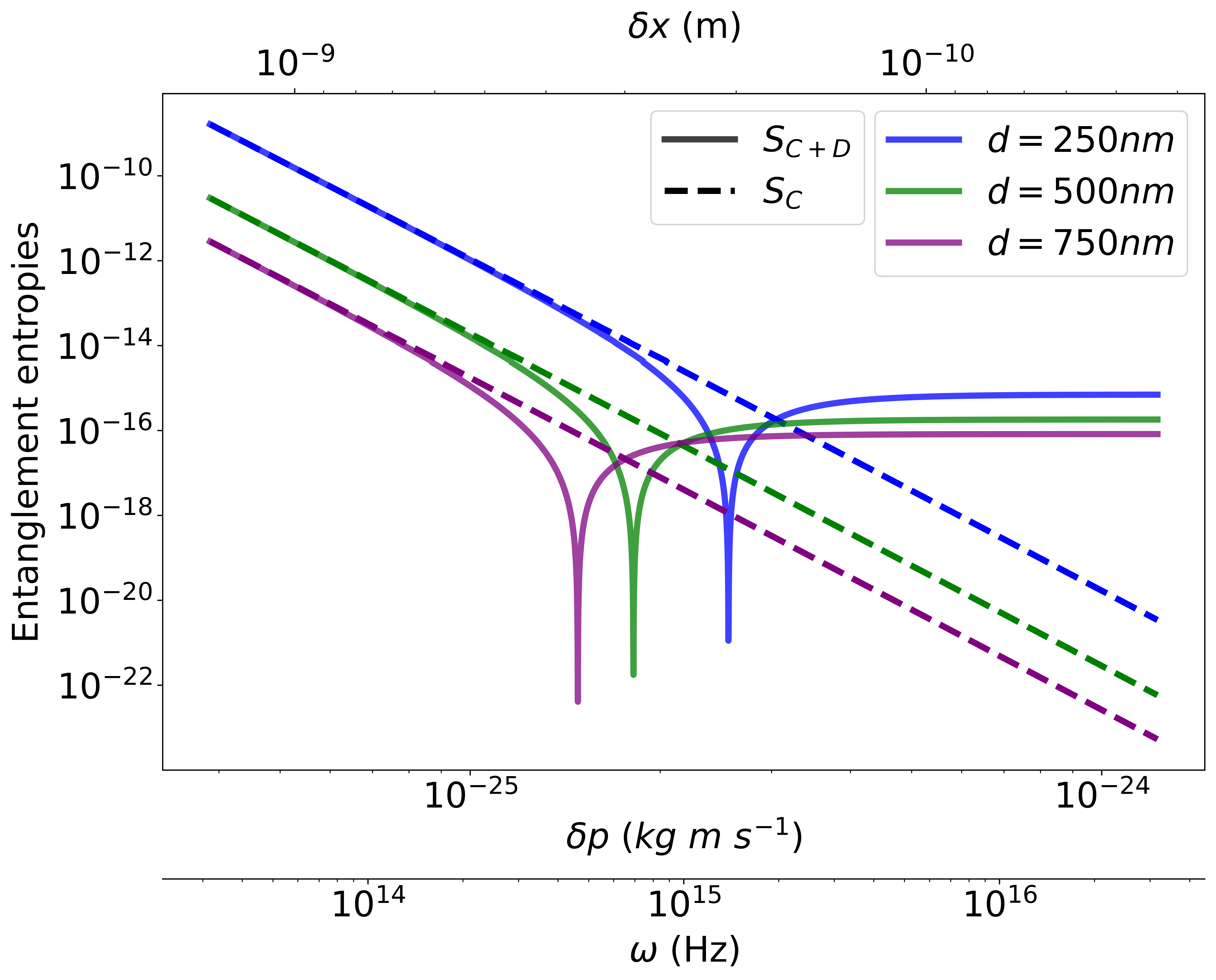}
    \caption{Comparison between entanglement entropies with and without relativistic effects considered.    
    The horizontal axis below the figure indicates the harmonic frequency, $\omega$. In contrast, the bottom (top) axis of the frame denotes the corresponding momentum (spatial) delocalization size given by $\delta p = \sqrt{\hbar m \omega/2}$ ($\delta x = \sqrt{\hbar/(2m \omega)}$). Dashed lines correspond to entanglement entropy arising only from the Coulomb interaction without relativistic corrections ($S_\text{C}$). In contrast, the solid lines correspond to the entanglement entropy, including the Darwin term ($S_\text{C+D}$). The inclusion of relativistic effects induces a frequency-independent entanglement background at high frequencies, which is graphically represented by the plateaus to the right side of the plot. Including relativistic corrections, the entanglement entropy vanishes at the points $\omega = 2 c/(\sqrt{3}d)$, which can be interpreted as a classicalization of the system (i.e., quantum correlations become vanishingly small). Similar ``dips'' also appear in the context of gravity~\cite{toros2024relativistic}, suggesting that such effects could be ubiquitous in nature, albeit in experimentally challenging regimes.
    \label{TimeIndependentEntropies}
    }
\end{figure}

In Fig.~\ref{TimeIndependentEntropies} we show that the the Darwin contribution becomes significant when we consider trapping frequencies around the petahertz regime for traps separated by nanoscale distances. The Darwin interaction induces a residual amount of entanglement at high frequencies, which appears as the non-zero plateau at high frequencies (while the contribution from the Coulomb interaction becomes vanishingly small in comparison). In addition, the 0PC Coulomb and 1PC Darwin contributions to the entanglement entropies cancel out when the mechanical frequency is tuned to the value $\omega = 2 c/(\sqrt{3}d)$, resulting in the entanglement dips. These nontrivial features, however, lie beyond the regime of existing experiments because of the high-frequency range and because it presents entanglement entropies of values below $10^{-14}$. At this point, we may consider which resources could be used to amplify the presence of these effects. In the next section, we show how the framework of continuous variable systems allows us to exploit squeezing to distinguish 0PC and 1PC effects in a more accessible regime.

\section{Dynamical Darwin Effects}\label{sec:darwin-dynamic}
In this section, we explore the interplay of Coulomb and Darwin effects in entanglement generation using squeezed states, which offer the possibility of controlling the initial delocalization in position and momentum in the context of Gaussian states~\cite{RevModPhys.84.621,serafini2003symplectic}. Our initial state will here consist of a product state of two single-mode squeezed vacuum states,
\begin{align} \label{init}
    \ket{\psi_0} = \ket{\xi}_1 \ket{\xi}_2,
\end{align}
where $\xi$ is the squeezing coefficient. The  single-mode squeezed vacuum state can be expressed in a number basis as: 
\begin{align}\label{heisenberg state}
    \ket{\xi} \equiv \left(\cosh{\xi}\right)^{\nicefrac{-1}{2}} \sum_{n=0}^{\infty} \left(\tanh{\xi}\right)^n \frac{\sqrt{(2n)!}}{2^{2n}} \ket{2n},
\end{align}
where $\ket{2n}$ denote even number states, and the delocalization of such states is given by
\begin{align} \label{DeltaxDeltap}
    \Delta x = \delta x\cdot e^{-\xi} \ ,  \ \ \Delta p= \delta p\cdot e^{\xi}.
\end{align}
The distance between the two traps, $d$, places a bound on the maximum position delocalization of each particle such that they remain separated in physical space. By imposing $\delta x \,e^{-\xi} \leq d$, and using Eq.~\eqref{zpf}, we obtain the bound: 
\begin{equation} \label{xix}
    \xi \geq \ln\sqrt{\frac{\hbar}{2m\omega d^2}} \equiv \xi_x.
\end{equation}
 We also obtain a constrain by noting that an initially delocalized state in momentum, i.e., with momentum delocalization $\delta p \,e^{\xi}$, will become delocalized in position, with position delocalization $\delta x \,e^{\xi}$, after a quarter rotation in phase space. Requiring $\delta x \, e^{\xi} \leq d$, and using Eq.~\eqref{zpf}, we now find the upper bound: 
\begin{equation}\label{xip}
    \xi \leq  \ln \sqrt{\frac{2m\omega d^2}{\hbar}} \equiv \xi_p.
\end{equation}
In other words, Eqs.~\eqref{xix} and ~\eqref{xip} constrain the position delocalization at $\omega t=0$ and at $\omega t =\pi/2$, respectively, where $\omega$ is the mechanical frequency and $t$ denotes the time. As we have $\xi_x = - \xi_p$, we find that the two bounds can be combined into the inequality $\vert \xi \vert \leq \xi_p$, which constrains the absolute value of the squeezing parameter~\footnote{We may consider some values as examples: using the frequency $\omega=1$GHz and distances $d$ of order $250$nm, $500$nm and $750$nm, we find that the squeezing parameter is constrained to the values $\vert \xi \vert \leq 0.06$, $\vert \xi \vert \leq 0.75$ and $\vert \xi \vert \leq 1.16$, respectively.
\label{footnote:squeezing}}.

The generated entanglement, and hence also any entanglement measure, will be a function of time. To quantify the evolution of the von Neumann entanglement entropy, we first need to solve the dynamics. The time evolution of the ladder operators is given by \cite{canosa2015exact}
\begin{align}
    \hat{a}_1(t) = k_0(t) \hat{a}_1 + k_{-}(t) \hat{a}_2 +k_{+}(t) \hat{a}_2^\dagger, \label{solution1}\\
    \hat{a}_2(t) = k_0(t) \hat{a}_2 - k_{-}(t) \hat{a}_1 - k_{+}(t) \hat{a}_1^\dagger,\label{solution2}
\end{align}
where $\hat{a}_{i}\equiv \hat{a}_{i}(t=0)$, and the coefficients $k_0$, $k_+$, $k_-$ are written as
\begin{align}
    k_0(t) &= \cos(\omega_\text{eff}t) - \frac{i\omega}{\omega_\text{eff}}\sin(\omega_\text{eff}t), \\
    k_{\pm}(t) &= (g_\text{C}\pm g_\text{D})\frac{\omega}{\omega_\text{eff}}\sin(\omega_\text{eff}t),
\end{align}
with $g_\text{C}$ and $g_\text{D}$ being the coupling constants defined in Eq. (\ref{couplings}). The effective frequency $\omega_\text{eff}$ is given by
\begin{align}\label{effective frequency}
    \omega_\text{eff} \equiv \sqrt{\omega^2 + (g_\text{C} - g_\text{D})^2 - (g_\text{C} + g_\text{D})^2},
\end{align}
which we note is a 1PC correction. Specifically, inserting the couplings from Eq.~\eqref{couplings} in Eq.~\eqref{effective frequency}, we find:
\begin{align}\label{effective frequency 2}
    \omega_\text{eff} = \omega \sqrt{ 1+  \frac{3 e^4}{16\pi^2 \varepsilon_0^2 m^2 \omega^2 c^2 d^4}}\,.
\end{align}
The dynamics of the system suffer a squeezing-independent frequency shift that appears only when relativistic effects are considered, since $\lim_{c\rightarrow \infty}\omega_\text{eff}=\omega$. 
Another interesting situation occurs when the frequency $\omega$ of the individual traps becomes small. In particular, in the limit of a vanishing trapping frequency $\omega$ we find
$\omega_\text{eff} \approx  \frac{\sqrt{3}e^2}{4\pi \varepsilon_0 m c d^2}$,
which shows that the relativistic effects can become the dominant contribution to the effective trapping frequency for certain parameter regimes.

Using Eqs.~\eqref{solution1}-\eqref{effective frequency 2}, we are now in a position to compute the time evolution of any expectation value. In particular, we will use them to compute the time-evolution of the von Neumann entanglement entropy $S(t)$. Since our Hamiltonian in Eq.~\eqref{hamiltonianequation} contains only quadratic operators, the evolution preserves the Gaussianity~\cite{walschaers2021non}. In particular, for Gaussian states, it is convenient to employ the covariance matrix formalism \cite{serafini2017quantum}, which allows an efficient way to compute the von Neumann entanglement entropy. As shown below, the computation reduces to the evaluation of time-dependent expectation values. The correspondence between density operators and covariance matrices can be found in \cite{case2008wigner,kubo1964wigner} and a summary on how to compute the von Neumann entanglement entropy is shown in Appendix~\ref{app:B}~\cite{demarie2012pedagogical,landi,RevModPhys.77.513,Reb_n_2011}. 

In this context, the von Neumann entanglement entropy is given by
\begin{align}\label{time-dependent entropy}
    S(t) &= - \sigma^{(1)}(t)\log \sigma^{(1)}(t) \nonumber\\
     &+ \big[1+\sigma^{(1)}(t)\big]\log \big[1+\sigma^{(1)}(t)\big],
\end{align}
where the symplectic eigenvalue $\sigma^{(1)}$ is 
\begin{align} \label{eq:second}
    \sigma^{(1)}(t) = \frac{1}{\hbar}\sqrt{4\langle  \hat{x}_i^2 (t)\rangle \langle  \hat{p}_i^2 (t)\rangle - \langle \{ \hat{x}_i (t), \hat{p}_i (t)\} \rangle^2 } - \frac{1}{2}.
\end{align}
The expectation values in Eq.~\eqref{eq:second} are evaluated using the state defined in Eq.~\eqref{heisenberg state}; the relation of the position and momentum operators to ladder operators is given in Eq.~\eqref{ladder} and the evolution of the ladder operators is provided in Eqs.~\eqref{solution1}-\eqref{effective frequency 2}. Using these latter relations, we eventually find:
\begin{align}\label{relevant eigenvalue}
    &\sigma^{(1)}(t)  =- \frac{1}{2}+\frac{\sqrt{2}}{4\omega_\text{eff}^2}\cdot\Bigg\{3e^{-4\xi}\left(e^{8\xi} g_\text{D}^2 + g_\text{C}^2\right)\omega^2 + 2\omega^4\nonumber\\
    &\,\,-4g_\text{D} g_\text{C} \left(g_\text{D}^2 - 6g_\text{D} g_\text{C} + g_\text{C}^2\right) +   
    \big(g_\text{D}^2 -8g_\text{D}g_\text{C} + g_\text{C}^2 \big)\omega^2 \nonumber\\
    &\,\, + e^{-4\xi } \Bigg[ -4 \big( g_\text{C}- e^{4\xi} g_\text{D}  \big)^2 \omega^2 \cos \left(2\omega_\text{eff}t\right)  \nonumber\\
    &\,\,\quad\quad\quad\quad  +\big[e^{8\xi}g_\text{D}^2\omega^2 + g_\text{C}^2 \omega^2 + e^{4\xi} \big(4g_\text{D}g_\text{C} (g_\text{D} + g_\text{C})^2 \nonumber\\
    &\,\,\quad\quad\quad\quad   -(g_\text{D}^2+g_\text{C}^2)\omega^2\big) \big]\cos\left(4\omega_\text{eff} t\right)\Bigg]\Bigg\}^{\nicefrac{1}{2}}.
\end{align}
Eqs.~\eqref{time-dependent entropy}  and ~\eqref{relevant eigenvalue} determine how the 0PC Coulomb and 1PC Darwin couplings affect the generation of the von Neumann entanglement entropy. In the following, we focus on investigating the parameter regimes of frequencies, distances and squeezings for which relativistic and non-relativistic cases could be distinguished, i.e. cases in which the entanglement entropies $S_C(t)$ and $S_{C+D}(t)$ could be discerned from one another. For this analysis, we constrain our squeezing parameter $\xi$ according to Eqs.~\eqref{xix} and \eqref{xip}.

\begin{figure*}
\centering
\raisebox{5mm}{\includegraphics[scale=0.085]{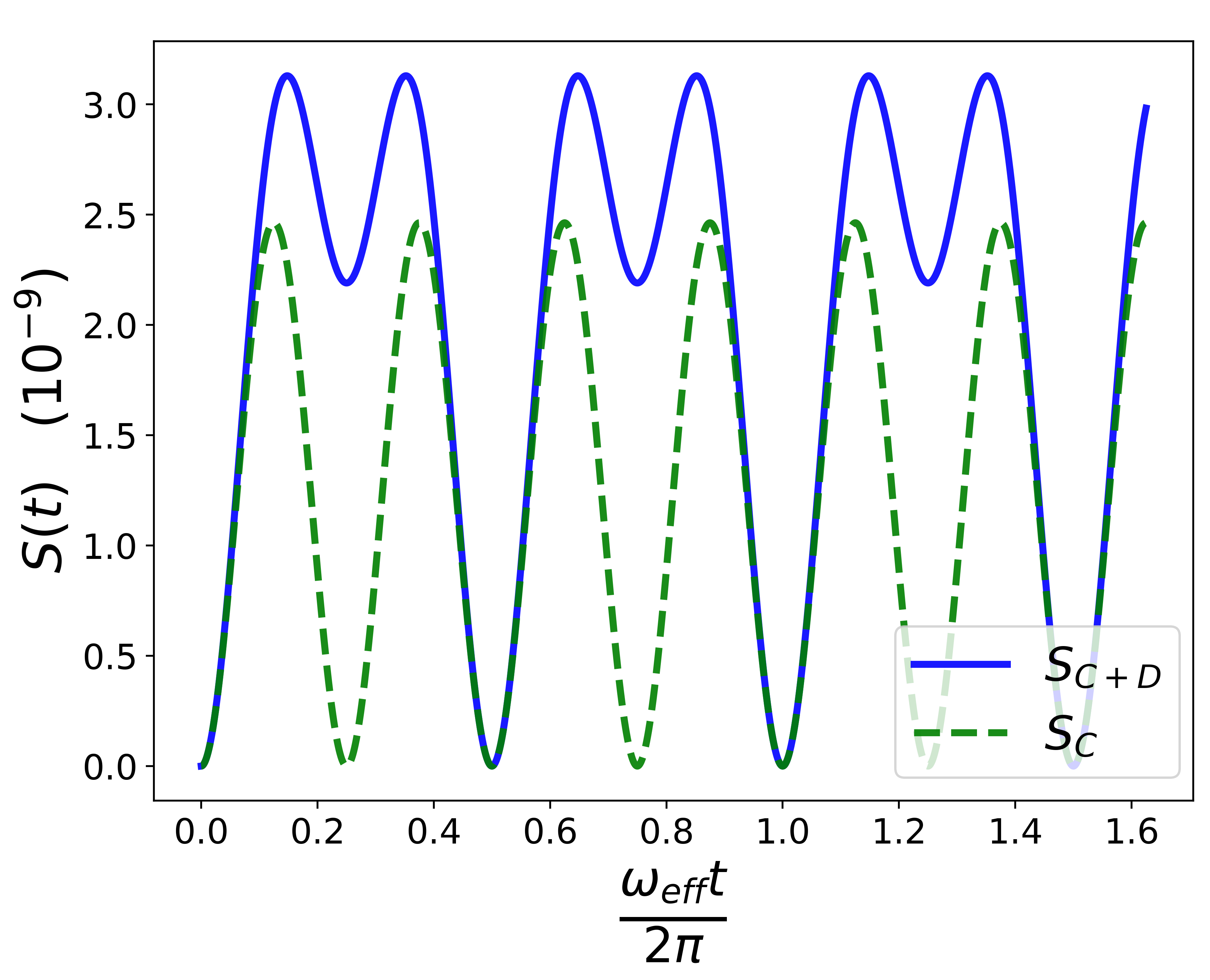}} 
\hspace{0.0cm}
\raisebox{5.5mm}{\includegraphics[scale=0.1275]{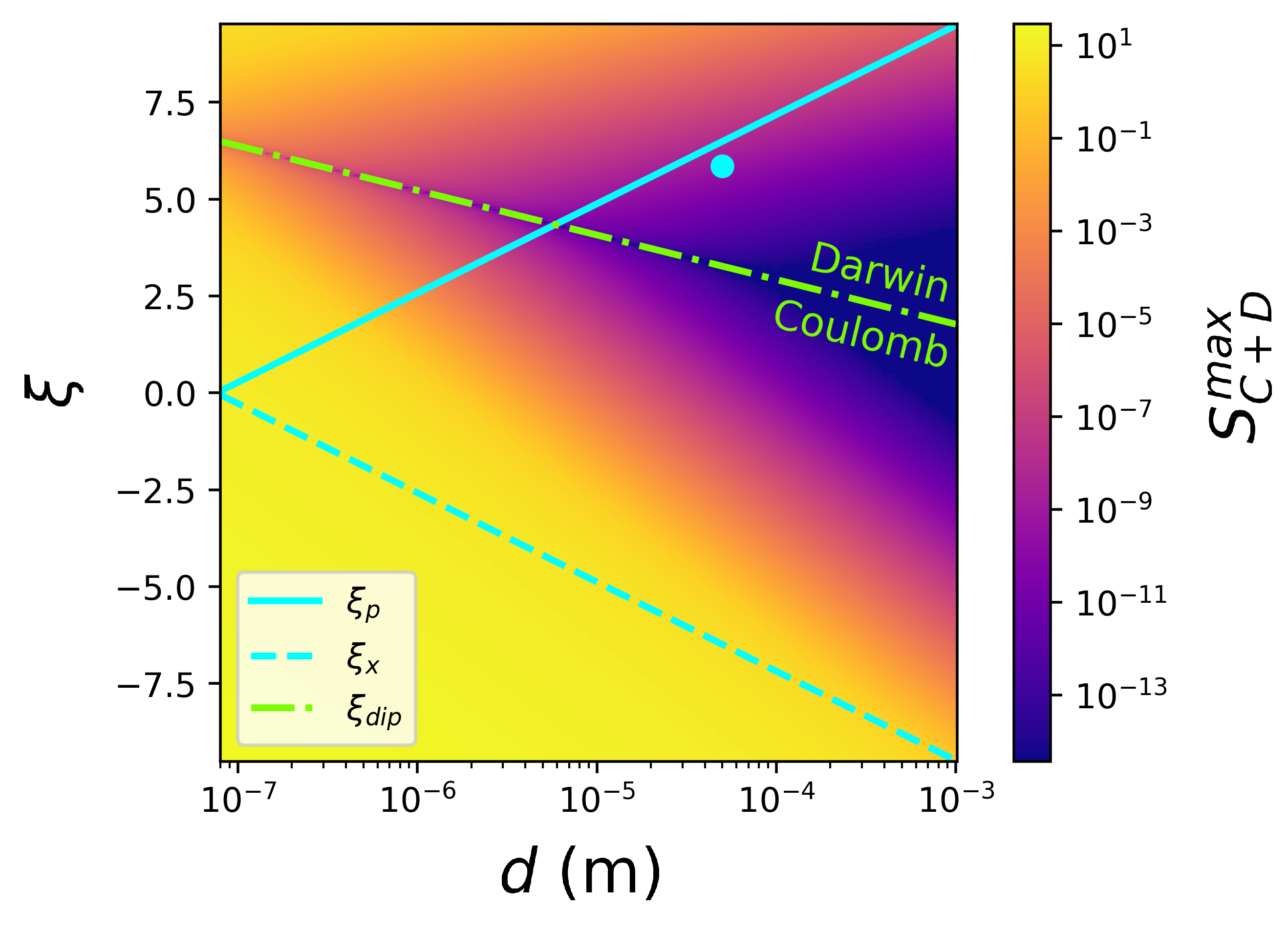}} 
\caption{
{\bf(left)} Comparison between time-dependent entanglement entropies with relativistic corrections ($S_\text{C+D}$) and without the relativistic corrections ($S_\text{C}$) during approximately $1$ns. The parameters used here are $\omega = 10$GHz, a distance of $d=50\mu$m and the squeezing $\xi = 6.16$, which is $95\%$ of the value allowed by the bound $\xi_p$ (see Eq.~\eqref{xip}). This figure indicates that there are regimes in which the relativistic contribution may dominate over non-relativistic interaction (i.e. including the Darwin term can increase the entanglement entropy) for certain frequencies and distances, even at short times.
{\bf(right)}
This figure shows the function $S^\text{max}_\text{C+D}\equiv \Max{t \in [0, \Delta t]}[S_\text{C+D}(t)]$ - that is, the maximum value achieved by the entanglement entropy in the relativistic case (considering Coulomb and Darwin couplings).
The frequency in this plot is fixed at $\omega = 10$GHz. The x-axis shows the distance between the traps ($d$), which ranges from approximately $0.1\mu$m to $1000\mu m$, while the y-axis shows the range of the squeezing parameter ($\xi$). The cyan lines correspond to the bounds defined by Eq.~(\ref{xix}) (dashed) and Eq.~(\ref{xip}) (solid), which were obtained by requiring that the two systems remain well separated in space; the allowed parameter space is located between the two lines. The maximum value achieved by the entanglement measure $S_\text{C+D}$ is higher in the yellowish parameter region and approaches zero in the darker regions, showing that there's a ``dip'' (i.e., vanishing of quantum correlations) highlighted by the green curve (Eq. (\ref{dip squeezing})), dividing the parameter space into two regions. Above (below) the green line, the Darwin contribution is bigger (smaller) than the Coulomb contribution to the generation of entanglement. The cyan dot represents the point whose parameters were used in the left figure.
} 
\label{Squeezing} 
\end{figure*}

As Eq. (\ref{relevant eigenvalue}) suggests, the time-dependent entropies have an oscillatory nature with a non-trivial dependence on the couplings and the squeezing parameter. One feature which distinguishes the relativistic and non-relativistic cases is the phase difference between the oscillations, which appears due to the difference between the frequencies $\omega$ and $\omega_\text{eff}$, the first associated with the non-relativistic dynamics and the latter appearing when both Coulomb and Darwin terms are considered. Another signature of relativistic effects appears when we compare the amplitude of $S_\text{C}(t)$ with $S_\text{C+D}(t)$. In Fig.~\ref{Squeezing} (left), we give an example in which the amplitudes of the two cases are distinct in that the Darwin term enhances the amount of entanglement between the traps. To explore more generally the parameter space of the theory, in Fig.~\ref{Squeezing} (right) we show the behaviour of the amplitude of the entanglement measure $S_\text{C+D}$ as a function of different values of squeezings and distances, using a fixed value for the frequency.


%





We observe that the amplitude of the von Neumann entropy $S_\text{C+D}(t)$ almost vanishes in the darker regions of the plot in Fig.~\ref{Squeezing} (right).
Such a dip in entanglement entropy, i.e., classicalization, occurs only when relativistic effects are considered, and it does not appear if we consider only $S_\text{C}(t)$. The found suppression of entanglement generation can be seen as the time-dependent analogue of the effect shown for the static case in Fig. \ref{TimeIndependentEntropies}.

The points in which the classicalization happens can be precisely determined by observing that the symplectic eigenvalue given in Eq. (\ref{relevant eigenvalue}) is maximal when we set the time to $t_\text{max} \equiv \frac{\pi}{2\omega_\text{eff}}$. Since $S_\text{C+D}$ is a strictly monotonically increasing function of $\sigma$, then  $S_\text{C+D}$ is also maximal at $t_\text{max}$. By imposing $S_\text{C+D}(t_\text{max})=0$, i.e., the generation of entanglement is fully suppressed, we find the following condition for the relativistic dip~\cite{toros2024relativistic}:
\begin{align}\label{dip condition}
    g_\text{C} e^{-2\xi} = g_\text{D} e^{2\xi}.
\end{align}
Eq.~\eqref{dip condition} tells us that the classicalization region is the region of the parameter space for which the entanglement entropy vanishes due to the competition of the Darwin and Coulomb couplings enhanced and diminished by the squeezing parameter, respectively. Using the definitions for $g_\text{C}$ and $g_\text{D}$ (see Eq. (\ref{couplings})), we find that the condition in Eq.~\eqref{dip condition} can be rewritten as,
\begin{align}\label{dip squeezing}
    \xi_\text{dip} = \frac{1}{4}\ln\left(\frac{4c^2}{3\omega^2 d^2}\right),
\end{align}
which was used for plotting the green curve in Fig.~\ref{Squeezing} (right). Eq.~\eqref{dip squeezing} delimits where each term (Darwin or Coulomb) is the dominant contribution in the generation of entanglement. The Coulomb interaction is dominant below the green curve, while above this threshold, the relativistic effects enhance the amount of entanglement between the two systems. We note that in the non-relativistic limit $c\rightarrow \infty$, the classicalization region does not exist, as there are no solutions of the non-relativistic limit of Eq.~\eqref{dip condition}, i.e., Eq.~\eqref{dip squeezing} diverges in the non-relativistic limit. In other words, the discovered entanglement dip is a genuinely relativistic effect.

The frequency used in Fig.~\ref{Squeezing} (right), $\omega=10$GHz, is in the range of current experiments~\cite{yu2022feasibility,yamahata2014gigahertz,glasson2005trapping,matthiesen2021trapping}, and this plot indicates possible parameters for probing the existence of the Darwin contribution in the entanglement generation. The available parameter space is constrained to the region between the solid and dashed cyan lines determined by the requirement that the two systems remain well separated in space (see Eqs.~\eqref{xix} and \eqref{xip}). The region in which the Darwin term dominates over the Coulomb term exists in a challenging region of the squeezing parameter, which is not achievable in current experiments (e.g., the cyan point indicated in the plot corresponding to  Fig.~\ref{Squeezing} (left)). However, we note that one could also probe the Darwin term below the threshold indicated by the green line, where the Coulomb term dominates, using precision measurements.


It is important to also consider possible sources of decoherence that may degrade the generated entanglement. One unavoidable source of decoherence for such setups is given by the emission of dipole radiation. The emission of radiation by electrons has been studied extensively in the literature (see the damping to the particle due to the Abraham-Lorentz-Dirac force~\cite{jackson1999classical}). The corresponding decoherence rate can be estimated as $\gamma = \frac{\mu_0 e^2 \omega^2}{6\pi m c}$~\cite{dirac1938classical,barut1980electrodynamics,rohrlich2000self} . Using the frequency $\omega = 10$ GHz and considering $m$ as the mass of an electron, the decoherence rate is $\gamma\approx 6.26\cdot 10^{-4} $Hz. There are however also other sources of decoherence for which we would need to compute the decoherence rate, e.g., ~\cite{yu2022feasibility,Fragolino:2023agd,Schut:2023tce}. To observe the generation of entanglement we would require that the total decoherence rate from all sources, $\gamma_\text{tot}$, is smaller than the inverse of the experimental time, $t_\text{tot}^{-1}$. In this paper, we will not delve into the details of various decoherence rates, and we will analyze separately the feasibility of the experiment and how to the witness the generated entanglement.


\section{Concluding Remarks}\label{sec:conclusion}

In this article, we explored how relativistic effects can influence the generation of entanglement in a system composed of two single-electron harmonic traps. Relativistic effects were included in the system's dynamics with the Darwin Hamiltonian, a 
post-Coulombian (PC) correction to the usual electrostatic potential.
We found that, at a quantum level, the Darwin contribution leads to a modifications in the build up of quantum correlations. 
Using the framework of continuous variables, we investigated cases where the entanglement is time-independent via perturbation theory and time-dependent by solving the dynamics. 

The static (time-independent)  case shows a frequency-independent entanglement background and ``dips'' representing the loss of quantum correlations, a sort of classicalization which also appears in the context of gravity ~\cite{toros2024relativistic}. Both features appear only for parameters beyond the scope of existing experiments. On the other hand, the dynamical (time-dependent) case can be more sensitive to relativistic corrections, and we provide values of frequencies and distances which future experiments with single-electron traps could explore. 
Also in this case we find regions where the Coulomb and Darwin contributions dominate in the generation of entanglement, as well as a region in-between, where the quantum correlations can become vanishingly small. The next task will be to study the experimental feasibility of our proposal and the development of a protocol for witnessing the generation of entanglement at the Coulomb-Darwin interface. \\

\begin{acknowledgements}

We want to thank Rajibul Islam for the discussions.
MT would like to acknowledge funding from ST/W006227/1. PA acknowledges funding from the Brazilian National Council for the Improvement of Higher Education (CAPES), Erasmus+ Programme, the Strategic Partnership Framework at the University of Groningen and Instituto Serrapilheira Call N$^\text{o}$4/2020. MS is supported by the Fundamentals of the Universe research program at the University of Groningen. SB thanks EPSRC grants EP/R029075/1, EP/X009467/1, and ST/W006227/1. SB and AM's research is supported by the Sloan, and Betty and Moore foundations. 
\end{acknowledgements}

\bibliographystyle{unsrt}
\bibliography{sample.bib}
\appendix
\section{Higher Order Darwin Quantum Corrections}\label{app:A}

In this Appendix, we provide higher-order corrections for the Darwin Hamiltonian. As we will see, higher orders of $\hbar$ will not maintain our interaction quadratic, so the analysis performed with the covariance matrix should suffer modifications considering higher order effects.

Let's denote the difference of the spatial fluctuation of the electrons as $\hat{\Delta} \equiv \hat{x}_2 - \hat{x}_1$. We will also retain only terms that generate entanglement, i.e. those that don't have identities acting on any party. The global Hamiltonian assumes the form
\begin{widetext}
\begin{align}\label{higher orders}
    \hat{H}_\text{int} &\approx \frac{e^2}{4\pi \varepsilon_0 d}\cdot  \Bigg[ 1+\left(1+\frac{\hat{\Delta}}{d}\right)  +\left(1+\frac{\hat{\Delta}}{d}\right)^2 \Bigg]\cdot \Bigg[ 1- \frac{\left({\hat{p}}_1 \cdot {\hat{p}}_2\right)}{2m^2c^2}\Bigg] \\
    &= \frac{e^2}{4\pi \varepsilon_0 d}\cdot \left[ \frac{\hat{\Delta}^2}{d^2} - \frac{3\left({\hat{p}}_1 \cdot {\hat{p}}_2\right)}{2m^2 c^2} - \frac{3}{2dm^2 c^2}\cdot \frac{\left\{ \hat{\Delta}, \left({\hat{p}}_1 \cdot {\hat{p}}_2\right)\right\}}{2} - \frac{1}{2d^2m^2c^2}\cdot \frac{\left\{\hat{\Delta}^2, \left({\hat{p}}_1 \cdot {\hat{p}}_2\right)\right\}}{2}\right].
\end{align}
\end{widetext}
The first term ($\propto \hat{\Delta}^2$) has order $\mathcal{O}(\hbar^1)$ and contains the standard Coulomb entangling bilinear $\delta \hat{x}_1\delta \hat{x}_2$. The second term ($\propto \left({\hat{p}}_1 \cdot {\hat{p}}_2\right)$) is the first order Darwin correction (also of order $\mathcal{O}(\hbar^{1})$). The third and fourth have, respectively, orders $\mathcal{O}(\hbar^{\nicefrac{3}{2}})$ and $\mathcal{O}(\hbar^{2})$. An ordering ambiguity arises from the notation of the classical derivation of the Darwin term at higher orders, so we symmetrize these terms via anticommutators to ensure hermiticity.

Assuming, for simplicity, that our state is initialized as $\ket{00}$, in this particular case, the Hamiltonian reads
\begin{widetext}
\begin{align}
\begin{aligned}
    \hat{H}_\text{int} &= \frac{\hbar e^2}{4\pi \varepsilon_0 d}\cdot \Bigg\{-\frac{1}{m\omega d^2}(\hat{a}_1^\dagger \hat{a}_2^\dagger) + \frac{3\omega}{4mc^2}(\hat{a}_1^\dagger \hat{a}_2^\dagger) - \frac{3}{2dc^2}\sqrt{\frac{\omega \hbar}{2m^3}}\left[2\left(\hat{a}_1^\dagger \hat{a}_2^{\dagger 2}-\hat{a}_1^{\dagger 2}\hat{a}_2^\dagger\right) +\left(\hat{a}_1 \hat{a}_1^\dagger \hat{a}_2^\dagger - \hat{a}_1^\dagger \hat{a}_2 \hat{a}_2^\dagger\right)\right] - \\
    &\hspace{3.625cm}-\frac{\hbar}{8d^2 m^2 c^2}\left[\left(\hat{a}_1^{\dagger 3}\hat{a}_2^\dagger + \hat{a}_1^\dagger \hat{a}_2^{\dagger 3}\right) - 2\left(\hat{a}_1^{\dagger 2}\hat{a}_2^{\dagger 2} - \hat{a}_1^{\dagger 2}\hat{a}_2 \hat{a}_2^\dagger- \hat{a}_1 \hat{a}_1^\dagger \hat{a}_2^{\dagger2} + \hat{a}_1 \hat{a}_1^\dagger \hat{a}_2 \hat{a}_2^\dagger\right)\right]\Bigg\}.
\end{aligned}
\end{align}
\end{widetext}
Neglecting, for simplicity, the set of operators which induce degeneracies to the ground state, according to the procedure detailed in Section \ref{sec:darwin-static}, the corresponding perturbed state is given by
\begin{widetext}
\begin{align}
\begin{aligned}\label{big_state}
    \ket{\psi} &= \frac{1}{\sqrt{\mathcal{N}}} \Bigg\{\ket{00} + \lambda_1 \ket{11} + \lambda_2 \left[2\left(\frac{\ket{12}-\ket{21}}{3}\right)+\left(\ket{01}-\ket{10}\right)\right] + \lambda_3 \Bigg[ \frac{(\ket{13}+\ket{31})}{4} +(\ket{20} + \ket{02}) + \frac{\ket{22}}{2} \Bigg]\Bigg\}, 
\end{aligned}
\end{align}
\end{widetext}
with the constants $\lambda_{1,2,3}$ having the following expressions
\begin{gather}
    \lambda_1 = \frac{e^2}{8\pi \varepsilon_0 md} \left[\frac{1}{(d\omega)^2}-\frac{3}{4c^2}\right], \\
    \lambda_2 = \frac{3e^2}{8\pi \varepsilon d^2 c^2}\sqrt{\frac{\hbar}{2\omega m^3}}, \\
    \lambda_3 = \frac{\hbar e^2}{32\pi \varepsilon_0 \omega m^2 c^2 d^3}.
\end{gather}
The order of magnitude of $\lambda_{2,3}$ suggests that their effects are too small to be detected in the usual frequency range used in experiments. Their inversely proportional dependence on the frequency shows that they tend to zero in the high-frequency regime, so the frequency-independent entanglement background seen in Fig. (\ref{TimeIndependentEntropies}) is an exclusively first-order effect and doesn't appear to be enhanced due to higher order corrections. The ``dips'' representing the loss of correlations - which appear in the relativistic cases presented in Fig. (\ref{TimeIndependentEntropies})) - will be slightly attenuated, in the sense of never being exactly zero, because $\lambda_{2,3}$ doesn't vanish in the points that $\lambda_1$ do. We express the density matrix associated with the state given in Eq (\ref{big_state}) for completeness.
\begin{widetext}
 \setcounter{MaxMatrixCols}{16}
\begin{align}
    \rho = \frac{1}{\mathcal{N}}
    \begin{pmatrix}
    1 & \lambda_2 & \lambda_3 & 0 & -\lambda_2 & \lambda_1 & \nicefrac{2\lambda_2}{3} & \nicefrac{\lambda_3}{4} & \lambda_3 & \nicefrac{-2\lambda_2}{3} & \nicefrac{\lambda_3}{4} & 0 & 0 & \nicefrac{\lambda_3}{4} & 0 & 0 \\
    \lambda_2 & \lambda_2^2 & \lambda_2 \lambda_3 & 0 & -\lambda_2^2 & \lambda_1\lambda_2 & \nicefrac{2\lambda_2^2}{3} & \nicefrac{\lambda_2\lambda_3}{4} & \lambda_2\lambda_3 & \nicefrac{-\lambda_2^2}{3} & \nicefrac{\lambda_2\lambda_3}{4} & 0 & 0 & \nicefrac{\lambda_2\lambda_3}{4} & 0 & 0 \\
    \lambda_3 & \lambda_2\lambda_3 & \lambda_3^2 & 0 & -\lambda_2\lambda_3 & \lambda_1\lambda_3  & \nicefrac{2\lambda_2\lambda_3}{3} & \nicefrac{2\lambda_3^2}{4} & \lambda_3^2 & \nicefrac{-2\lambda_2\lambda_3}{3} & \nicefrac{\lambda_3^2}{4} & 0 & 0 & \nicefrac{\lambda_3^2}{4} & 0 & 0 \\
    0 & 0 & 0 & 0 & 0 & 0 & 0 & 0 & 0 & 0 & 0 & 0 & 0 & 0 & 0 & 0 \\
    -\lambda_2 & - \lambda_2^2 & -\lambda_2\lambda_3 & 0 & \lambda_2^2 & -\lambda_1 \lambda_2 & \nicefrac{-2\lambda_2^2}{3} & \nicefrac{-\lambda_2\lambda_3}{4} & -\lambda_2\lambda_3 & \nicefrac{2\lambda_2^3}{3} & \nicefrac{-\lambda_2\lambda_3}{4} & 0 & 0 & \nicefrac{-\lambda_2\lambda_3}{4} & 0 & 0 \\
    \lambda_1 & \lambda_1\lambda_2 & \lambda_1\lambda_3 & 0 & -\lambda_1\lambda_2 & \lambda_1^2 & \nicefrac{2\lambda_1\lambda_2}{3} & \nicefrac{\lambda_1\lambda_3}{4} & \lambda_1\lambda_3 & \nicefrac{-2\lambda_1 \lambda_2}{3} & \nicefrac{\lambda_1\lambda_3}{4} & 0 & 0 & \nicefrac{\lambda_1\lambda_3}{4} & 0 & 0\\
    \nicefrac{2\lambda_2}{3} & \nicefrac{2\lambda_2^2}{3} & \nicefrac{2\lambda_2\lambda_3}{3} & 0 & \nicefrac{-\lambda_2^2}{3} & \nicefrac{2\lambda_1\lambda_2}{3} & \nicefrac{4\lambda_2^2}{9} & \nicefrac{\lambda_2\lambda_3}{6} & \nicefrac{2\lambda_2\lambda_3}{3} & \nicefrac{-4\lambda_2^2}{9} & \nicefrac{\lambda_2\lambda_3}{6} & 0 & 0 & \nicefrac{\lambda_2\lambda_3}{6} & 0 & 0  \\
    \nicefrac{\lambda_3}{4} & \nicefrac{\lambda_2\lambda_3}{4} & \nicefrac{\lambda_3^2}{4} & 0 & \nicefrac{-\lambda_2\lambda_3}{4} & \nicefrac{\lambda_1\lambda_3}{4} & \nicefrac{\lambda_2\lambda_3}{6} & \nicefrac{\lambda_3^2}{16} & \nicefrac{\lambda_3^2}{4} & \nicefrac{-\lambda_2\lambda_3}{6} & \nicefrac{\lambda_3^2}{16} & 0 & 0 & \nicefrac{\lambda_3^2}{16} & 0 & 0 \\
    \lambda_3 & \lambda_2\lambda_3 & \lambda_3^2 & 0 & -\lambda_2\lambda_3 & \lambda_1\lambda_3 & \nicefrac{2\lambda_2\lambda_3}{3} & \nicefrac{\lambda_3^2}{4} & \lambda_3^2 & \nicefrac{-2\lambda_2\lambda_3}{3} & \nicefrac{\lambda_3^2}{4} & 0 & 0 & \nicefrac{\lambda_3^2}{4} & 0 & 0  \\
    \nicefrac{-2\lambda_2}{3} & \nicefrac{-\lambda_2^2}{3} & \nicefrac{-\lambda_2\lambda_3}{3} & 0 & \nicefrac{\lambda_2^2}{3} & \nicefrac{-2\lambda_1 \lambda_2}{3} & \nicefrac{-4\lambda_2^2}{9} &\nicefrac{-\lambda_2\lambda_3}{6} & \nicefrac{-2\lambda_2\lambda_3}{3} & \nicefrac{4\lambda_2^2}{9} & \nicefrac{-\lambda_2 \lambda_3}{6} & 0 & 0 & \nicefrac{-\lambda_2 \lambda_3}{6} & 0 & 0 \\
    \nicefrac{\lambda_3}{2} & \nicefrac{\lambda_2\lambda_3}{2} & \nicefrac{\lambda_3^2}{2} & 0 & \nicefrac{-\lambda_2 \lambda_3}{2} & \nicefrac{\lambda_1\lambda_3}{2} & \nicefrac{\lambda_2 \lambda_3}{3} & \nicefrac{\lambda_3^2}{8} & \nicefrac{\lambda_3^2}{2} & \nicefrac{-\lambda_2\lambda_3}{3} & \nicefrac{\lambda_3^2}{8} & 0 & 0 & \nicefrac{-\lambda_2\lambda_3}{6} & 0 & 0  \\
    0 & 0 & 0 & 0 & 0 & 0 & 0 & 0 & 0 & 0 & 0 & 0 & 0 & 0 & 0 & 0 \\
    0 & 0 & 0 & 0 & 0 & 0 & 0 & 0 & 0 & 0 & 0 & 0 & 0 & 0 & 0 & 0 \\
    \nicefrac{\lambda_3}{4} & \nicefrac{\lambda_2\lambda_3}{4} & \nicefrac{\lambda^2_3}{4} & 0 & \nicefrac{-\lambda_2\lambda_3}{4} & \nicefrac{\lambda_1\lambda_3}{4} & \nicefrac{\lambda_2\lambda_3}{6}  & \nicefrac{\lambda_3^2}{16} & \nicefrac{\lambda_3^2}{4} & \nicefrac{-\lambda_2\lambda_3}{6} & \nicefrac{\lambda_3^2}{16} & 0 & 0 & \nicefrac{\lambda_3^2}{16} & 0 & 0 \\
    0 & 0 & 0 & 0 & 0 & 0 & 0 & 0 & 0 & 0 & 0 & 0 & 0 & 0 & 0 & 0 \\
    0 & 0 & 0 & 0 & 0 & 0 & 0 & 0 & 0 & 0 & 0 & 0 & 0 & 0 & 0 & 0 \\
    \end{pmatrix}
\end{align}
\end{widetext}

\section{Computing the von Neumann entanglement entropy for Gaussian states}\label{app:B}
In this Appendix, we provide a short review of the computation of the von Neumann entanglement using covariance matrices \cite{demarie2012pedagogical,landi,RevModPhys.77.513}), which is a useful method for continuous variable systems.

As our system is Gaussian, instead of the usual density operator, we may employ the formalism of covariance matrices in order to characterize the correlations between the two subsystems. Using the positions and momenta written in terms of the ladder operators (Eq. (\ref{ladder})), we define the quadrature vector
\begin{align}
    \boldsymbol{Y} \equiv (\hat{x}_1, \hat{p}_1, \hat{x}_2, \hat{p}_2),
\end{align}
such that the usual commutation relations can be written as
\begin{align}
    [\boldsymbol{Y}_j, \boldsymbol{Y}_k] = 2i\boldsymbol{\Omega}_{jk},
\end{align}
where $\boldsymbol{\Omega}$ is known as the symplectic metric, which in our bipartite case is given by \cite{serafini2017quantum,landi,demarie2012pedagogical}
\begin{align}
    \boldsymbol{\Omega} = \bigoplus^{2}_{j=1} (i \sigma_y) = 
    \begin{pmatrix}
    0 & 1 & 0 & 0 \\
    -1 & 0 & 0 & 0 \\
    0 & 0 & 0 & 1 \\
    0 & 0 & -1 & 0
    \end{pmatrix}.
\end{align}
The global covariance matrix is defined with the covariances of the components of $\boldsymbol{Y}$, i.e.
\begin{align}
    \sigma_{ij} \equiv \langle \{ Y_i, Y_j \}\rangle - 2 \langle Y_i \rangle \langle Y_j \rangle,
\end{align}
and for our system, the covariance matrix is explicitly given by
\begin{align}
\boldsymbol{\sigma}=
\begin{pmatrix}
2\left\langle \hat{x}_1^2\right\rangle & \left\langle\left\{\hat{x}_1, \hat{p}_1\right\}\right\rangle & 2\left\langle \hat{x}_1 \hat{x}_2\right\rangle & 2\left\langle \hat{x}_1 \hat{p}_2\right\rangle \\
\left\langle\left\{\hat{x}_1, \hat{p}_1\right\}\right\rangle & 2 \left\langle \hat{p}_1^2\right\rangle & 2\left\langle \hat{p}_1 \hat{x}_2\right\rangle & 2\left\langle \hat{p}_1 \hat{p}_2\right\rangle \\
2\left\langle \hat{x}_2 \hat{x}_1\right\rangle & 2\left\langle \hat{x}_2 \hat{p}_1\right\rangle & 2\left\langle \hat{x}_2^2\right\rangle &\left\langle\left\{\hat{x}_2, \hat{p}_2\right\}\right\rangle \\
2\left\langle \hat{p}_2 \hat{x}_1\right\rangle & 2\left\langle \hat{p}_2 \hat{p}_1\right\rangle & \left\langle\left\{\hat{x}_2, \hat{p}_2\right\}\right\rangle & 2\left\langle \hat{p}_2^2\right\rangle
\end{pmatrix}.
\end{align}
Together with the symplectic metric, the covariance matrix satisfies the positive definite relation 
\begin{align}
\boldsymbol{\sigma}+i \boldsymbol{\Omega} \geq 0,
\end{align}
which is a generalized way to write the Robertson-Schrödinger uncertainty relation, a more general uncertainty inequality than Heisenberg's principle \cite{landi}. The reduced covariance matrices can be extracted from the diagonal blocks,
\begin{align}\label{covariance matrix}
    \boldsymbol{\sigma}_i = 
    \begin{pmatrix}
    2\langle \hat{x}_i^2 \rangle & \langle \{ \hat{p}_i,  \hat{x}_i \} \rangle \\
    \langle \{ \hat{p}_i, \hat{x}_i \} \rangle & 2\langle \hat{p}_i^2 \rangle 
    \end{pmatrix} , \ \ i = 1,2.
\end{align}

As the entanglement entropy of the system can be evaluated with the von Neumann entropy of the reduced state of the first subsystem, $S = -\text{Tr}(\rho_1 \log \rho_1)$, we can also characterise this entanglement measure in terms of the covariance matrix of the first party. Since we assume that two electrons are equally coupled (i.e. their traps have equal frequencies), the reduced covariance matrices are isospectral.

From Eq.~\eqref{hamiltonianequation}, the entanglement entropy is determined by the symplectic eigenvalues 
of the reduced covariance matrix, $\sigma^{(1)}$ and $\sigma^{(2)}$. The symplectic eigenvalues are different from the usual eigenvalues of an operator, and they're obtained by calculating the spectrum of $\boldsymbol{\sigma}\boldsymbol{\Omega}$. The symplectic spectrum $\{\sigma^{(i)}\}$ is invariant under symplectic transformations and determines the properties of the Gaussian states. The expressions for the symplectic eigenvalues are  \cite{Reb_n_2011}
\begin{align}\label{eigenvalues}
    \sigma^{(1)} = \frac{1}{\hbar}\sqrt{\det\left(\boldsymbol{\sigma}_1\right)} - \frac{1}{2} \ , \ \ \sigma^{(2)} = - 1 - \sigma^{(1)}.
\end{align}
In terms of the elements of the reduced covariance matrix, the symplectic eigenvalue $\sigma^{(1)}$ - which can be interpreted as the effective occupation number of the mode \cite{Reb_n_2011} - is given by Eq.~\eqref{eq:second}. Using this quantity, the entanglement entropy is given by Eq.~\eqref{time-dependent entropy}. 

An important feature of our model that allows us to use this formalism easily is that our Hamiltonian is quadratic so that the Gaussianity of the covariance matrix is preserved during the evolution of the system - which also implies that each reduced covariance matrix remains Gaussian \cite{Canosa_2020,Reb_n_2011}. Gaussianity is not necessarily preserved when higher-order corrections are considered, as the Hamiltonian is not quadratic.

An alternative and equivalent formula for parametrizing the entanglement entropy in terms of the covariance matrix is given by \cite{kumar2023continuous}
\begin{align}
    S = f\left(\frac{1}{\hbar}\sqrt{\det{\boldsymbol{\sigma}_1}}\right),
\end{align}
with the function $f$ being expressed as
\begin{align}
    f(x) & = \left( x+\frac{1}{2}\right)\log\left( x+\frac{1}{2}\right) - \left( x-\frac{1}{2}\right)\log\left( x+\frac{1}{2}\right).
\end{align}

\end{document}